\documentclass[conference,10pt]{IEEEtran}
\usepackage[utf8]{inputenc}
\usepackage{color,xcolor}
\usepackage{epsfig}
\usepackage{amsmath,amsfonts,amsthm,amssymb}
\usepackage{graphics,graphicx,pgfplots,tikz,subfigure}
\usepackage{float}
\usepackage{algorithm,algorithmic}
\usepackage{svg}
\usepackage{epstopdf}
\usepackage{soul,csquotes,ulem}

\newcommand{\eqdef}{\stackrel{\text{def}}{=}}
\newcommand{\figref}[1]{Fig. \ref{#1}}

\begin{document}
\title{An application-oriented scheduler}
	\author{
		\IEEEauthorblockN{
			J.-C. Sibel, N. Gresset, V. Corlay
		}
		\IEEEauthorblockA{
			Mitsubishi Electric R\&D Centre Europe\\
			Rennes, France\\
			Email: \{j.sibel, n.gresset, v.corlay\}@fr.merce.mee.com
		}
	}
	\maketitle
	
\begin{abstract}
	We consider a multi-agent system where agents compete for the access to the radio resource. 
	By combining some application-level parameters, such as the resilience, with a knowledge of the radio environment, we propose a new way of modeling the scheduling problem as an optimization problem. 
	We design accordingly a low-complexity solver. The performance are compared with state-of-the-art schedulers via simulations.
	The numerical results show that this application-oriented scheduler performs better than standard schedulers. 
	As a result, it offers more space for the selection of the application-level parameters to reach any arbitrary performance.
\end{abstract}


\section{Introduction\label{sec_intro}}
	In modern wireless communications, latency and reliability are as important as throughput. Indeed, new applications, such as the industrial internet of things, involve new use cases with very high latency and reliability requirements.  For instance, \cite{Baek2021} describes latency requirements of 0.5 ms and reliability of 99.999999$\%$ for motion control. Similar figures are provided for mobile automation. These requirements should be met in the context of systems with many agents sharing the same communication resources. Moreover, for agents with different missions, their application requirements and the quality of their channel may vary.

	In this scope, we believe that all available degrees of freedom should be used in the design of the communication system.
	More specifically, the application requirements should be taken into account in the optimization to maximize the performance. 
	Having this paradigm in mind, we shall focus on the critical MAC layer in this paper, namely the scheduling. 

	Of course, there exist many studies in the literature that propose design guidelines with respect to these new requirements. 
	A relevant emerging field is the Age of Information (AoI).
	AoI measures the time elapsed between the generation of a message $u(t)$ and its delivery time $t$, i.e., the AoI is $t-u(t)$. 
	This metric enables to assess existing queuing and scheduling strategies and can serve as a design guideline for new algorithms.
	AoI is a finer metric than the quality of service (QoS), used for resource allocation, e.g. in the LTE \cite{ref_quali}, as the latter is transmission centric while the former focuses on the quality of the information effectively received\footnote{Note that the QoS could also be considered as a problem variable in an AoI optimization problem}.
	AoI is also different from resource allocation mechanisms considering communication metrics, such as the data rate or the channel capacity, for their scheduling decision~\cite{Gresset2015}. 
	However, the standard AoI metric does not take into account application requirements. It is possible to weight the AoI of each agent (in the case where a sum AoI is optimized) but this does not enable to directly optimize with respect to these requirements. Moreover, according to many definitions a \textit{real-time} constraint means meeting a deadline. This is slightly different than the metric optimized with AoI. 

	\textbf{Related work -} In \cite{Kaul2012}, the average time status update with the queue algorithm first-come-first-served is investigated. 
	In \cite{Kaul2011}, the same authors study AoI in the context of a vehicular network. It is shown that the source rate should not be too high to maximize the AoI metric.
	Moreover, several scheduling algorithms have been proposed to optimize AoI: the AoI in the case of multiple agents, each having multiple sources, and sharing a common channel is considered in \cite{He2018}. 
	The authors prove that the problem is NP-hard and proposed sub-optimal algorithms.
	In \cite{Kadota2018}, the case of a base-station delivering information to several agents is studied. 
	The context is similar to the one in this paper, but the optimization is done with respect to a different metric: they investigate a transmission scheduling policy that minimizes the expected weighted sum of the AoI of each agent. 
	
	\textbf{Main contributions -} In this work, we define a new multi-agent scheduling problem. It takes several application parameters into account, including the resilience, i.e., how often an application needs to receive fresh information to work properly. This enables to assess the quality of a scheduler with respect to the application failure probability. As a result, similarly to what is done with semantic communications \cite{Weng2021}, the communication system is optimized directly with respect to its final real-time requirements. We show with numerical evaluations a strong confidence in the proposed scheduler design as the performance gain is significant compared with state-of-art schedulers.	 

\section{Context}
	\subsection{Description of the system}
\begin{figure}[!h]
			\centering
			\includegraphics[width=\linewidth]{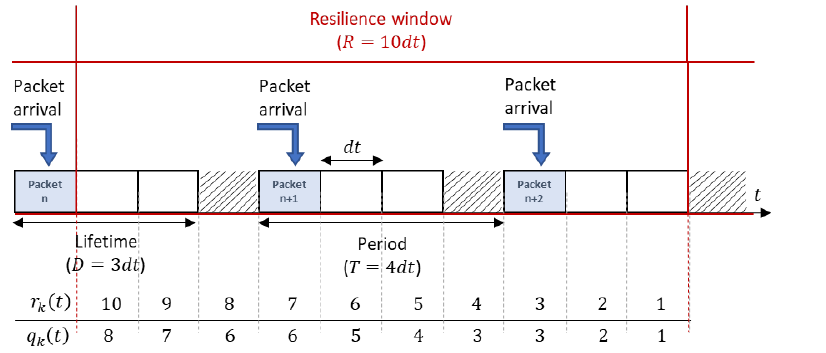}
			\caption{Example of a time-line with $T=4$ and $D=3$ (in number of $dt$'s).}
			\label{fig:Timeline}
		\end{figure}

		We consider a discrete-time system, divided in time slots whose length is denoted by $dt$, e.g., $dt=1$ msec. Let $N$ be the number of agents in the system. Any $k^{th}$ agent $A_k$ is represented by a data stream characterized by three parameters: 
		\begin{itemize}
			\item The period $T$: the time duration between two consecutive packet arrivals from the application in the agent buffer for transmission. 
			\item The packet lifetime $D$ (with $D \leq T$): the time duration for which the packet is alive.
			\item The resilience $R$: the maximum time duration for which the agent can survive without a successful transmission of a packet. 
		\end{itemize}
		\figref{fig:Timeline} shows a representation, called \textit{time-line},  of $T$ and $D$. \\
	
		We define the event \textit{resilience violation} for any agent (E1) as \enquote{no packet successfully transmitted during the last $R$ time slots}. For an agent $A_k$ at (discrete) time $t$, we accordingly introduce $r_k(t)$ as the remaining time before resilience violation. This means that $0 \leq r_k(t) \leq R$ and $r_k(t)$ is a decreasing function of $t$. The complementary event of (E1), called \textit{success} (E0), is: \enquote{successful transmission of a packet at time $t$}. With both the events (E0) and (E1) for any agent $A_k$ at time $t$, $r_k(t)$ is immediately set to its maximum value $R$. 
		
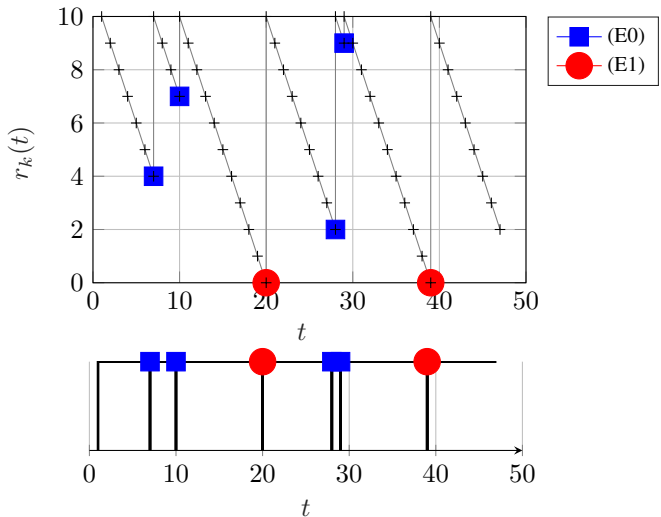
\begin{figure}[!h]
			\centering
%
%
\begin{tikzpicture}

\begin{axis}[%
	width = 0.65\linewidth,
	height=0.15\textheight,
	scale only axis,
	xmin=0,
	xmax=50,
	xlabel style={font=\color{white!15!black}, anchor=south west, at={(0.45,-0.1)}},
	xlabel={$t$},
	ymin=0,
	ymax=10,
	ylabel style={font=\color{white!15!black}, anchor=south west, at={(0.1,0.35)}},
	ylabel={$r_{k}(t)$},
	axis background/.style={fill=white},
	xmajorgrids,
	ymajorgrids,
	legend style={legend cell align=left, align=left, draw=white!15!black, at={(1.05,1)}, anchor=north west, font=\footnotesize}
]
\addplot [color=blue, draw=none, mark size=3.5pt, mark=square*, mark options={solid, fill=blue, blue}]
  table[row sep=crcr]{%
7	4\\
10	7\\
28	2\\
29	9\\
};
\addlegendentry{(E0)}

\addplot [color=red, draw=none, mark size=5.0pt, mark=*, mark options={solid, fill=red, red}]
  table[row sep=crcr]{%
20	0\\
39	0\\
};
\addlegendentry{(E1)}

\addplot [color=gray, mark size=2pt, mark=+, mark options={black}, forget plot]
  table[row sep=crcr]{%
1	10\\
2	9\\
3	8\\
4	7\\
5	6\\
6	5\\
7	4\\
7	10\\
8	9\\
9	8\\
10	7\\
10	10\\
11	9\\
12	8\\
13	7\\
14	6\\
15	5\\
16	4\\
17	3\\
18	2\\
19	1\\
20	0\\
20	10\\
21	9\\
22	8\\
23	7\\
24	6\\
25	5\\
26	4\\
27	3\\
28	2\\
28	10\\
29	9\\
29	10\\
30	9\\
31	8\\
32	7\\
33	6\\
34	5\\
35	4\\
36	3\\
37	2\\
38	1\\
39	0\\
39	10\\
40	9\\
41	8\\
42	7\\
43	6\\
44	5\\
45	4\\
46	3\\
47	2\\
};
\end{axis}
\end{tikzpicture}%
\\
			\hspace*{-1.5em}
%
%
\begin{tikzpicture}

\begin{axis}[%
	width = 0.65\linewidth,
	height=0.05\textheight,
	scale only axis,
	xmin=0,
	xmax=50,
	xlabel style={font=\color{white!15!black}},
	xlabel={$t$},
	ymin=0,
	ymax=1,
	axis background/.style={fill=white},
	axis lines=left, ytick=\empty,
	y axis line style={draw=none},
	xmajorgrids,
]

\addplot [color=blue, draw=none, mark size=3.5pt, mark=square*, mark options={solid, fill=blue, blue}]
  table[row sep=crcr]{%
7	1\\
10	1\\
28	1\\
29	1\\
};

\addplot [color=red, draw=none, mark size=5pt, mark=*, mark options={solid, fill=red, red}]
  table[row sep=crcr]{%
20	1\\
39	1\\
};

\addplot [color=black, line width=1pt, forget plot]
  table[row sep=crcr]{%
1	0\\
1	1\\
7	1\\
7	0 \\
7	1\\
10	1\\
10	0\\
10	1\\
20	1\\
20	0\\
20	1\\
28	1\\
28	0\\
28	1\\
29	1\\
29	0\\
29	1\\
39	1\\
39	0\\
39	1\\
47	1\\
};
\end{axis}
\end{tikzpicture}%
			\caption{Example of a time evolution of $r_k(t)$ with $R=10$ (top) and the associated resilience windows (bottom). At time instants $t=7,11,31,33$, $A_k$ meets (E0) whereas at time instants $t=22,44$, $A_k$ meets (E1).}
			\label{fig:EvolutionOf_rk}
		\end{figure}

		We define a resilience window as a time window that starts just after one event (E0) or one event (E1), i.e., when $r_k(t)=R$, and that ends when $A_k$ meets either the next (E0) or the next (E1). \figref{fig:EvolutionOf_rk} shows an arbitrary example of the time evolution of $r_k(t)$ and the associated resilience windows.
	
		Additionally, we define a transmission opportunity for an agent $A_k$ as a time slot for which a packet is alive. Accordingly, we define the quantity $q_k(t)$ as the remaining number of transmission opportunities before meeting (E1). In case $D=T$, i.e., if there is no \enquote{hole} in the time-line, then $q_k(t)=r_k(t)$. In case $D < T$, then $q_k(t) \leq r_k(t)$. 
	
	\subsection{Description of the environment}
		We consider a single available radio resource per time slot. We assume that the resource is suited to the transmission of one packet by any agent. Therefore, at any time, all the agents with an alive packet compete for the access to the single radio resource but only one agent finally obtain the resource. We denote by $\delta_k(t)=1$ the event \enquote{$A_k$ is allocated at time $t$} and by $\delta_k(t)=0$ the event \enquote{$A_k$ is not allocated at time $t$}. When allocated the resource, $A_k$ observes an unsuccessful transmission of the packet with probability $p_k$ being the channel error probability. Thus, an agent $A_k$ meets (E1) either if it is never allocated the resource or if the channel strongly impacts the transmission each time the agent is allocated.
		
	
	\subsection{Example\label{subsec:Example}}
		Consider a mobile robot with a trajectory monitoring application: $T$ is the time duration between two consecutive position measurements of the agent, $D$ is the time duration for which the measured position remains relevant, and $R$ represents the capacity to interpolate/extrapolate the trajectory without consecutive measurements. We consider that the agent's mission is to reach a geographical point. 
		
		Assume that due to numerous clutters in its surrounding environment, some packets are dropped by the channel resulting in event (E1) for the agent. The trajectory cannot be interpolated/extrapolated with a sufficient accuracy to ensure the success of the mission. Therefore, the agent stops its motion to wait for a full restart which generates, among others, a significant delay. Such an event is then detrimental to the system performance regarding the application purpose. 
		
		In our study, for simplification purpose, the restart delay is not considered. In other words, the event (E1) does not stop the agent.

\section{Optimization problem\label{sec:Optimization}}
	
	In this section, we first formalize the scheduling problem as an optimization problem. 	
	Then, we introduce some heuristics to allow for a practical solution.

	\subsection{Presentation of the problem}
		The problem we propose to solve is the opportunistic centralized scheduling problem: \enquote{which agent, at a given time slot, should get the radio ressource given the knowledge of the system and the environment?}. 
		For any agent $A_k$ we define $V_k(t)$ as the accumulated sum of resilience violations until time $t$. $V_k(t)$ is updated based on the events (E0) and (E1) as follows:
		\begin{align}
			\text{Meet (E0):}\quad & V_k(t) \leftarrow V_k(t-dt),\\
			\text{Meet (E1):}\quad & V_k(t) \leftarrow V_k(t-dt) + 1.
		\end{align}
		If none of (E0) or (E1) occurs at time $t$, then $V_k(t)$ is naturally extended as $V_k(t)\leftarrow V_k(t-dt)$. From these quantities, we define the local \textit{long-term} \underline{experimented} probability of resilience violation at any time $t$ as:
		\begin{equation}
			F_k(t) \eqdef \frac{V_k(t)}{t}.
			\label{eq:Fk}
		\end{equation}
		From a system perspective, we define the optimization problem as the search for the scheduling decision $\underline{\delta}^*(t) = \{\delta_1^*(t),...,\delta_N^*(t)\}$ that minimizes the average probability of resilience violation $F(t)$ embodied by the sum of $F_k$'s s.t.:
		\begin{equation}
			F(t) \eqdef \frac{\sum_k V_k(t)}{t}.
			\label{eq:PerfMetric}
		\end{equation}
		
	\subsection{Proposed alternative objective\label{sec:ProposedAlternativeObjective}}	
		As a first step to design an efficient scheduler, we build a heuristic $J\bigl(t,\underline{\delta}(t)\bigr)$ whose aim is to predict $F(t)$ by predicting actually:
		\begin{equation}
			S(t) \eqdef \sum_k V_k(t)
			\label{eq:S(t)}
		\end{equation}	
		under a given scheduling decision $\underline{\delta}(t)$. The rationale behind considering $S(t)$ instead of $F(t)$ is that $t$ is common to all the agents therefore $t$ only scales the problem, i.e., there is no need to insert it within the optimization problem. 
		
		The solver in the following section then consists in choosing the allocation $\underline{\delta}(t)$ such that the heuristic $J\bigl(t,\underline{\delta}(t)\bigr)$ is minimized. To establish the said heuristic, first, we build a function $\hat{V}\bigl(t,\delta_k(t)\bigr)$ to estimate $V_k(t)$ for any agent $A_k$. Then, the heuristic $J\bigl(t,\underline{\delta}(t)\bigr)$ is obtained by summing these local estimates.

		\subsubsection{Local predict function\label{sec:Optimization_Local}}
			As $V_k(t)$ is an observation metric, we define an associated local \textit{long-term} \underline{predicted} accumulated sum of resilience violations $\hat{V}_k\bigl(t,\delta_k(t)\bigr)$:
			\begin{equation}
				\hat{V}_k\bigl(t,\delta_k(t)\bigr) \eqdef V_k(t-dt) + \hat{f}_k\bigl(t,\delta_k(t)\bigr),
				\label{eq:LocalLongTermPrediction}
			\end{equation}
			where $V_k(t-dt)$ is the observed value from the previous time instant and $\hat{f}_k\bigl(t,\delta_k(t)\bigr)$ is the local \textit{short-term} \underline{predicted} probability of resilience violation within the current resilience window for time $t$ given an allocation decision $\delta_k(t)$. 
			The function $\hat{f}_k(\cdot,\cdot)$ acts as a correction term especially when $t$ is small. For example, when all $V_k$'s are zero at the beginning, they are not well-representing the near future. 
			
			We construct $\hat{f}_k(\cdot,\cdot)$ to fairly represent the current application status as well as the radio conditions:
			\begin{itemize}
				\item $\hat{f}_k(\cdot,\cdot)$ increases as $p_k$ increases: a worse radio conditions makes greater the probability of resilience violation 
				\item $\hat{f}_k(\cdot,\cdot)$ increases as $r_k(t)$ decreases: getting closer to the resilience violation makes greater the probability of resilience violation 
			\end{itemize}
			We propose the heuristic $f_r(\cdot,\cdot)$ that fulfills these requirements 
			\begin{equation}
				f_r(t,k) \eqdef p_k^{r_k(t)}.
			\end{equation}
			In the case of holes in the time-line, i.e., when $D<T$, two agents $A_{k_1},A_{k_2}$ with the same values $r_{k_1}(t) = r_{k_2}(t)$ can observe two different time-lines $q_{k_1}(t) < q_{k_2}(t)$. We propose the following enhancement of the heuristic to distinguish between $A_{k_1}$ and $A_{k_2}$:
			\begin{equation}
				f_q(t,k) \eqdef p_k^{q_k(t)}.
				\label{eq:ExpectedInstantaneousProbabilityOfFailure}
			\end{equation}		
			Now, we build the function $\hat{f}_k(\cdot,\cdot)$ to integrate any of these heuristics. The function should distinguish the case the allocation is not granted from the case the allocation is granted. In the latter case, indeed, $\hat{f}_k(\cdot,\cdot)$ must depend on $p_k$ because a channel transmission is assumed. As a result, we propose the following definition:
			\begin{equation}
				\hat{f}_k\bigl(t,\delta_k(t)\bigr) = \left\{
									\begin{array}{ll}
										f(t,k)     & \text{if } \delta_k(t)=0,\\
										p_k f(t,k) & \text{if } \delta_k(t)=1,
									\end{array}
								      \right.
			\end{equation}
			where $f$ could be either $f_r$ or $f_q$. This can be simplified as:
			\begin{equation}
				\hat{f}_k\bigl(t,\delta_k(t)\bigr) = \bigl(1-\delta_k(t)(1-p_k)\bigr)f(t,k).
				\label{eq:LocalPrediction}
			\end{equation}

			Now that $\hat{V}_k$ is fully constructed, the local heuristic is nearly completed. We use the utility-based formalism to define the local predict function:
			\begin{equation}
				j_k\bigl(t,\delta_k(t)\bigr) \eqdef U_{\alpha}\Bigl(\hat{V}_k\bigl(t,\delta_k(t)\bigr)\Bigr),
			\end{equation}
			where $U_{\alpha}(\cdot)$ is a non-linear monotonic function identical for all the agents.  
			In this paper, we consider the $\alpha$-fair utility function \cite{Lan2010}:
			\begin{equation}
				U_{\alpha}(x) \eqdef \left\{
							\begin{array}{ll}
								\frac{x^{1-\alpha}}{1-\alpha} & \alpha \neq 1,\\
								\log x & \alpha = 1
							\end{array}
						     \right., \quad U_{\alpha}'(x) = x^{-\alpha}.
				\label{eq:Utility}
			\end{equation}		
			The $\alpha$-fair utility framework allows us for considering a family of schedulers with a good performance/fairness tradeoff \cite{Schwarz2011}. The rationale behind the introduction of $U_{\alpha}$ is to stick with such a well-known family of schedulers. Several values of $\alpha$ are considered in this paper to observe if the fairness concern indeed exerts any influence on the scheduling performance. 
		\subsubsection{Global predict function\label{sec:Optimization_Local}}
					
			%
			We define the global predict function of $S(t)$ as the sum of the local predict functions:
			\begin{equation}
				J\bigl(t,\underline{\delta}(t)\bigr) \eqdef \sum_k j_k\bigl(t,\delta_k(t)\bigr).
				\label{eq:SumCost}
			\end{equation}
			The goal of the solver is then to find the scheduling decision $\underline{\delta}^*(t)$ that minimizes $J$:
			\begin{equation}
				\underline{\delta}^*(t) = \arg \min_{\underline{\delta}(t)} J\bigl(t,\underline{\delta}(t)\bigr).
				\label{eq:OptimizationProblem}
			\end{equation}
			
\section{Solver}
	In this section, we present two manners for solving \eqref{eq:OptimizationProblem}: the exact solver and the approximated solver. These two solvers are important regarding the computational complexity because a great number of agents leads the exact solver to overload any computation resource. 
	\subsection{Exact solver for $J$: On-line sum\label{sec:OLS}}
	%
		Let us first compute $j_k(t,1)$ and $j_k(t,0)$ for any agent $A_k$ having an alive packet at time $t$. This corresponds to the events \enquote{$A_k$ is allocated} and \enquote{$A_k$ is not allocated}, respectively. If $t$ is a time slot out of the lifetime packet of $A_k$, i.e., the packet of $A_k$ is dead, $A_k$ is not considered for the scheduling decision. $j_k(t,1)$ and $j_k(t,0)$ are set to infinite values to explicitly exclude $A_k$. This provides us Table \ref{tab:OnlineSum}. 
		\begin{table}[!h]
			\centering
			\begin{tabular}{|c||c|c|c|c|}
				\hline
				$k^*$ 		         &      $0$ &      $1$ & $\dots$  & $N-1$      \\ \hline\hline
				$j_1\bigl(t,\delta_1(t)\bigr)$     & $j_1(t,1)$ & $j_1(t,0)$ & $\dots$  & $j_1(t,0)$ \\ \hline
				$j_2\bigl(t,\delta_2(t)\bigr)$     & $j_2(t,0)$ & $j_2(t,1)$ & $\dots$  & $j_2(t,0)$ \\ \hline
				$\vdots$	         & $\vdots$ & $\vdots$ & $\vdots$ & $\vdots$ \\ \hline
				$j_N\bigl(t,\delta_N(t)\bigr)$     & $j_N(t,0)$ & $j_N(t,0)$ & $\dots$  & $j_N(t,1)$ \\ \hline
			\end{tabular}
			\vspace*{1em}
			\caption{Table of the on-line sum solver}
			\label{tab:OnlineSum}
		\end{table}
		
		Secondly, for each scheduler decision, i.e., for each column in Table \ref{tab:OnlineSum}, we sum all the rows to obtain a list of $N$ cost values. Thirdly, we extract the column index $k^*$ whose cost value is lower than any other cost value. This leads to consider agent $A_{k^*}$ as the agent to allocate. This solver requires at least $N$ times the computations of $j_k(t,1),j_k(t,0)$, then $N$ sums of $N$ terms each (so $N^2$ operations at least), then a comparison between $N$ values. Consequently, it might cause computational issues when $N$ increases.
	\subsection{Approximated solver for $J$: On-line Taylor\label{sec:OLT}}
		For great values of $N$, let us use the Taylor expansion of $U_{\alpha}(\cdot)$ around the prediction $\hat{V}_k\bigl(t,\delta_k(t)\bigr)$ considering the experimented $V_k(t-dt)$:
		\begin{eqnarray}
			U_{\alpha}\Bigl(\hat{V}_k\bigl(t,\delta_k(t)\bigr)\Bigr) & \approx & U_{\alpha}\bigl(V_k(t-dt)\bigr) \\
										   & +       & \Bigl(\hat{V}_k\bigl(t,\delta_k(t)\bigr)-V_k(t-dt)\Bigr) \nonumber\\
										   & \times  & U_{\alpha}'\bigl(V_k(t-dt)\bigr) \nonumber.
		\end{eqnarray}
		We only keep the terms that depend on the current scheduling decision at $t$. In addition, as only one agent is provided the resource at a time, this leads \eqref{eq:OptimizationProblem} to become:
		\begin{equation}
			k^* = \arg \min_k \sum_k \hat{V}_k\bigl(t,\delta_k(t)\bigr)U_{\alpha}'\bigl(V_k(t-dt)\bigr).
		\end{equation}
		First, we replace the differential from \eqref{eq:Utility}, then we replace the prediction from \eqref{eq:LocalLongTermPrediction} and finally we only keep what depends on the scheduling decision $\underline{\delta}(t)$. We get:
		\begin{equation}
			k^* = \arg \min_k \sum_k \hat{f}_k\bigl(t,\delta_k(t)\bigr) V_k(t-dt)^{-\alpha}.
		\end{equation}
		Using \eqref{eq:LocalPrediction} and keeping again only the terms that depend on the current scheduling decision $\underline{\delta}(t)$, the previous equation becomes:
		\begin{equation}
			k^* = \arg \max_k \sum_k \delta_k(t)(1-p_k)f(t,k) V_k(t-dt)^{-\alpha}.
		\end{equation}
		One agent $A_{k^*}$ is allocated at a time $t$, therefore, $\delta_{k^*}(t)=1$ whereas $\delta_{k\neq k^*}(t)=0$. Accordingly, the previous equations amounts to:
		\begin{eqnarray}
			k^* & = & \arg \max_k \left\{M_k(t)\right\}_k,\\
			M_k(t) & \eqdef &(1-p_k)f(t,k) V_k(t-dt)^{-\alpha} \nonumber.
		\end{eqnarray}
		The on-line Taylor solver requires a comparison between $N$ metrics $M_1(t),\dots,M_N(t)$, each one requiring few computations, which is dramatically less than what requires the on-line sum. For a large quantity of agents, therefore, provided that the Taylor expansion holds, i.e., $|\hat{V}_k\bigl(t,\delta_k(t)\bigr)-V_k(t-dt)| < \epsilon$ with $\epsilon \ll 1$, it is highly recommended to consider this solver.

\section{Numerical observations\label{sec:NumericalEvaluations}}
	This section presents the evaluation results of the proposed solution in comparison with other schedulers from the state-of-the-art. 
	
	\subsection{Challengers}
		We perform the On-Line Sum (OLS) scheduler from \ref{sec:OLS} and the On-Line Taylor (OLT) scheduler from \ref{sec:OLT} considering either $f=f_r$ (the schedulers are then called OLS-R and OLT-R, respectively) or $f=f_q$ (the schedulers are then called OLS-Q and OLT-Q, respectively) when $D=T$ and when $D<T$. We set the $\alpha$ parameter of the utility function $U_{\alpha}$ to $\alpha\in\{-5,-2,-1,0\}$, see \cite{Schwarz2011} for details on the fairness impact.
		
		We confront OLS and OLT to the Round-Robin scheduler \cite{ArpaciDusseau18-Book} used for network scheduling. It consists in allocating the radio resource to the agents one by one following a buffer. The said buffer is a random permutation of $[0,\dots,N-1]$. In case an agent $A_k$ does not have an alive packet at time $t$, i.e., the packet is dead, the scheduler scans the next buffer indexes to extract the first agent with an alive packet. In addition, when the Round-Robin has finished a round in its buffer, i.e., after $N$ allocation steps, the Round-Robin replaces its buffer with a new random permutation of $[0,\dots,N-1]$. This randomization prevents an agent from being always out at each period of the Round-Robin.
		
		We also confront the on-line scheduler to a proportional-fair like scheduler (PF-like)
		whose allocation rule is based on the channel capacity:
		\begin{equation}
			k^* = \arg \max_{k} \frac{\log_2\bigl(1+\gamma_k(t)\bigr)}{\sum_{t'=0}^{t-dt}\omega_k(t')\log_2\bigl(1+\gamma_k(t')\bigr)}.
		\end{equation}
		with $\gamma_k(t)$ the instantaneous signal-to-noise ratio at time $t$ (dual value of $p_k$) and $\omega_k(t')=1$ if $A_k$ had an alive packet at time $t'$ and $\omega_k(t')=0$ otherwise. The denominator comprises the accumulated quantity of resources allocated in the past to the agent in terms of the channel capacity and the numerator indicates the instantaneous capacity the agent can reach at time $t$. Therefore, with two agents with the same past, the scheduler allocates the agent with the greatest capacity. With two agents with the same instantaneous capacity, the scheduler allocates the agent whose accumulated capacity is the lowest. Therefore, PF-like balances between good performance and fairness.
		
	\subsection{Environment}
		We perform 1000 iterations of 10000 time slots duration each for each of the challengers. At each iteration, each probability $p_k$ is drawn randomly around a mean value $\bar{p}_k$. This actually creates a non-static environment for the agents. The mean values are linearly selected in $[10^{-3},\dots,10^{-1}]$ such that each agent $A_k$ is provided $\bar{p}_k \neq \bar{p}_{k'\neq k}$.
		
		We consider a drop of $N=100$ agents with a packet period $T=100$, with two lifetimes $D=90,100$ to observe the behavior difference by selecting $f=f_q$ and $f=f_r$. 
		
		We consider that the time-line of an agent does not necessarily starts at the same time slot as another agent. Therefore, at each iteration, the time start of each agent is randomly drawn between zero and $T$.	
	
	\subsection{Performance metric}
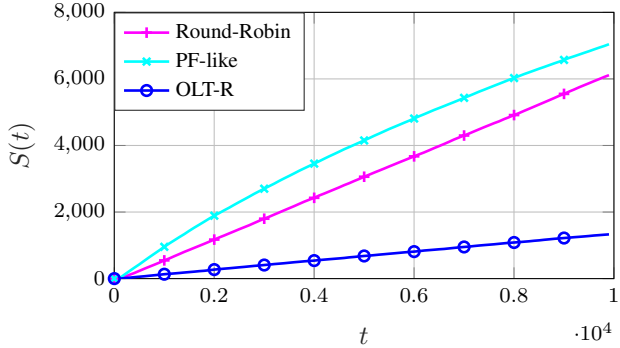
\begin{figure}[!h]
			\centering
%
%
\definecolor{mycolor1}{rgb}{1.00000,0.00000,1.00000}%
\definecolor{mycolor2}{rgb}{0.00000,1.00000,1.00000}%
\begin{tikzpicture}

\begin{axis}[%
	width=0.75\linewidth,
	height=0.15\textheight,
	font =\footnotesize,
	at={(0.758in,0.481in)},
	scale only axis,
	xmin=0,
	xmax=10000,
	xlabel style={font=\color{white!15!black}},
	xlabel={$t$},
	ymin=0,
	ymax=8000,
	ylabel style={font=\color{white!15!black}},
	ylabel={$S(t)$},
	axis background/.style={fill=white},
	xmajorgrids,
	ymajorgrids,
	legend style={legend cell align=left, align=left, draw=white!15!black, at={(0,1)}, anchor=north west}
]
\addplot [color=mycolor1, line width=1.0pt, mark size=2.0pt, mark=+, mark options={solid, mycolor1}, mark repeat = 10]
  table[row sep=crcr]{%
1	0\\
101	0\\
201	48\\
301	112\\
401	168\\
501	232\\
601	296\\
701	357\\
801	417\\
901	479\\
1001	547\\
1101	606\\
1201	673\\
1301	737\\
1401	803\\
1501	863\\
1601	924\\
1701	985\\
1801	1046\\
1901	1109\\
2001	1172\\
2101	1237\\
2201	1296\\
2301	1359\\
2401	1421\\
2501	1484\\
2601	1544\\
2701	1604\\
2801	1671\\
2901	1737\\
3001	1800\\
3101	1859\\
3201	1924\\
3301	1987\\
3401	2052\\
3501	2115\\
3601	2178\\
3701	2245\\
3801	2311\\
3901	2370\\
4001	2427\\
4101	2497\\
4201	2555\\
4301	2617\\
4401	2679\\
4501	2741\\
4601	2808\\
4701	2866\\
4801	2932\\
4901	2994\\
5001	3061\\
5101	3122\\
5201	3185\\
5301	3245\\
5401	3306\\
5501	3369\\
5601	3430\\
5701	3492\\
5801	3552\\
5901	3613\\
6001	3673\\
6101	3734\\
6201	3791\\
6301	3855\\
6401	3921\\
6501	3982\\
6601	4045\\
6701	4111\\
6801	4180\\
6901	4240\\
7001	4299\\
7101	4359\\
7201	4423\\
7301	4486\\
7401	4549\\
7501	4609\\
7601	4670\\
7701	4725\\
7801	4792\\
7901	4850\\
8001	4914\\
8101	4973\\
8201	5039\\
8301	5102\\
8401	5166\\
8501	5230\\
8601	5294\\
8701	5358\\
8801	5423\\
8901	5490\\
9001	5550\\
9101	5618\\
9201	5680\\
9301	5747\\
9401	5809\\
9501	5870\\
9601	5933\\
9701	5993\\
9801	6056\\
9901	6112\\
};
\addlegendentry{Round-Robin}

\addplot [color=mycolor2, line width=1.0pt, mark size=2.0pt, mark=x, mark options={solid, mycolor2}, mark repeat = 10]
  table[row sep=crcr]{%
1	0\\
101	7\\
201	117\\
301	223\\
401	327\\
501	427\\
601	536\\
701	643\\
801	746\\
901	849\\
1001	955\\
1101	1056\\
1201	1150\\
1301	1245\\
1401	1336\\
1501	1436\\
1601	1531\\
1701	1620\\
1801	1711\\
1901	1803\\
2001	1890\\
2101	1970\\
2201	2053\\
2301	2132\\
2401	2218\\
2501	2301\\
2601	2381\\
2701	2463\\
2801	2548\\
2901	2626\\
3001	2701\\
3101	2778\\
3201	2853\\
3301	2935\\
3401	3012\\
3501	3086\\
3601	3162\\
3701	3240\\
3801	3312\\
3901	3383\\
4001	3455\\
4101	3526\\
4201	3601\\
4301	3674\\
4401	3743\\
4501	3814\\
4601	3887\\
4701	3952\\
4801	4018\\
4901	4085\\
5001	4151\\
5101	4220\\
5201	4291\\
5301	4357\\
5401	4425\\
5501	4494\\
5601	4556\\
5701	4619\\
5801	4683\\
5901	4745\\
6001	4810\\
6101	4878\\
6201	4939\\
6301	5003\\
6401	5067\\
6501	5125\\
6601	5184\\
6701	5245\\
6801	5304\\
6901	5366\\
7001	5431\\
7101	5490\\
7201	5553\\
7301	5617\\
7401	5675\\
7501	5735\\
7601	5794\\
7701	5853\\
7801	5912\\
7901	5973\\
8001	6026\\
8101	6085\\
8201	6144\\
8301	6197\\
8401	6251\\
8501	6306\\
8601	6357\\
8701	6411\\
8801	6465\\
8901	6516\\
9001	6571\\
9101	6630\\
9201	6679\\
9301	6730\\
9401	6782\\
9501	6832\\
9601	6886\\
9701	6938\\
9801	6987\\
9901	7042\\
};
\addlegendentry{PF-like}

\addplot [color=blue, line width=1.0pt, mark size=2.0pt, mark=o, mark options={solid, blue}, mark repeat = 10]
  table[row sep=crcr]{%
1	0\\
101	0\\
201	12\\
301	27\\
401	43\\
501	57\\
601	70\\
701	86\\
801	101\\
901	116\\
1001	132\\
1101	144\\
1201	156\\
1301	169\\
1401	183\\
1501	196\\
1601	209\\
1701	224\\
1801	235\\
1901	252\\
2001	271\\
2101	285\\
2201	299\\
2301	311\\
2401	325\\
2501	339\\
2601	352\\
2701	367\\
2801	382\\
2901	394\\
3001	408\\
3101	421\\
3201	437\\
3301	449\\
3401	462\\
3501	477\\
3601	492\\
3701	505\\
3801	519\\
3901	530\\
4001	541\\
4101	557\\
4201	570\\
4301	583\\
4401	595\\
4501	607\\
4601	620\\
4701	638\\
4801	651\\
4901	666\\
5001	680\\
5101	691\\
5201	705\\
5301	720\\
5401	733\\
5501	747\\
5601	760\\
5701	774\\
5801	788\\
5901	800\\
6001	811\\
6101	826\\
6201	844\\
6301	856\\
6401	868\\
6501	881\\
6601	891\\
6701	907\\
6801	923\\
6901	938\\
7001	951\\
7101	962\\
7201	979\\
7301	994\\
7401	1007\\
7501	1019\\
7601	1031\\
7701	1046\\
7801	1061\\
7901	1075\\
8001	1085\\
8101	1096\\
8201	1109\\
8301	1121\\
8401	1137\\
8501	1149\\
8601	1166\\
8701	1180\\
8801	1195\\
8901	1207\\
9001	1219\\
9101	1231\\
9201	1244\\
9301	1255\\
9401	1267\\
9501	1280\\
9601	1295\\
9701	1306\\
9801	1318\\
9901	1329\\
};
\addlegendentry{OLT-R}

\end{axis}
\end{tikzpicture}%
			\caption{Accumulated sum of violations $S(t)$ over the time.}
			\label{fig:SumOfViolations}
		\end{figure}

		We consider $F(t)$ defined in \eqref{eq:PerfMetric} as the performance metric. First, let us observe the time evolution of $S(t)$ from \eqref{eq:S(t)} for some challengers, see \figref{fig:SumOfViolations}. The curves reach a linear steady state after around 5000 time slots whatever the scheduler. 
		
		\noindent
		$F(t)$ being the slope of $S(t)$, we conclude that $F(t)$ reaches a constant value $F_{\text{cst}}$ for a sufficiently large amount of time $t$. Practically speaking, we consider the performance metric to be $F_{\text{cst}}$ computed as the slope of the curve of $S(t)$ over the last 5000 time slots of the simulation. To ensure an even more reliable $F_{\text{cst}}$ value, we average over the previously mentioned 1000 iterations for each scheduler, for each value of $R$.
		
	\subsection{Results}
\begin{figure}[!h]
			\centering
%
%
\definecolor{mycolor1}{rgb}{1.00000,0.00000,1.00000}%
\definecolor{orange}{rgb}{1,0.6,0.2}%

\begin{tikzpicture}

\begin{axis}[%
	width=0.75\linewidth,
	height=0.25\textheight,
	font=\footnotesize,
	scale only axis,
	xmin=90,
	xmax=115,
	xlabel style={font=\color{white!15!black}},
	xlabel={$R$},
	ymode=log,
	ymin=0.00001,
	ymax=1,
	yminorticks=true,
	ylabel style={font=\color{white!15!black}},
	ylabel={$F_{\text{cst}}$},
	axis background/.style={fill=white},
	title style={font=\bfseries},
	axis x line*=bottom,
	axis y line*=left,
	xmajorgrids,
	ymajorgrids,
	yminorgrids,
	legend style={at={(0.97,0.03)}, anchor=south east, legend cell align=left, align=left, draw=white!15!black, at={(0,0)}, anchor=south west}
]

\addplot [color=black, dashed, line width=1.0pt]
  table[row sep=crcr]{%
90	0.62039217650744\\
95	0.559587085610065\\
100	0.505722438918934\\
105	0.457503089048855\\
110	0.411662240604426\\
115	0.368357860184376\\
};
\addlegendentry{Round-Robin}

\addplot [color=gray, line width=1.0pt]
  table[row sep=crcr]{%
90	0.589960147162806\\
95	0.488118505707031\\
100	0.3\\
105	0.211755354384941\\
110	0.162976885473875\\
115	0.16144110716039\\
};
\addlegendentry{PF-like}

\addplot [color=orange, line width=1.0pt, mark size=2.0pt, mark=+, mark options={solid, orange}]
  table[row sep=crcr]{%
90	0.13157174018647\\
95	0.0738452870658115\\
100	0.0216465299935788\\
105	0.00128249982296769\\
110	3.39461191760266e-06\\
115	0\\
};
\addlegendentry{OLS-R, $\alpha=0$}

\addplot [color=orange, dotted, line width=1.0pt, mark size=2.0pt, mark=+, mark options={solid, orange}]
  table[row sep=crcr]{%
90	0.13157174018647\\
95	0.0738560719645017\\
100	0.0216951326891731\\
105	0.00129371173450709\\
110	4.98076149446856e-06\\
115	0\\
};
\addlegendentry{OLT-R, $\alpha=0$}


\addplot [color=red, dotted, line width=1.0pt, mark size=2.0pt, mark=x, mark options={solid, red}]
  table[row sep=crcr]{%
90	0.13254798976832\\
95	0.0742295874973899\\
100	0.0217158848191035\\
105	0.00126533659663452\\
110	3.4805801392232e-06\\
115	0\\
};
\addlegendentry{OLT-R, $\alpha=-1$}

\addplot [color=green, dotted, line width=1.0pt, mark size=2.0pt, mark=o, mark options={solid, green}]
  table[row sep=crcr]{%
90	0.132473162594178\\
95	0.0743867314370222\\
100	0.0234739537714012\\
105	0.00127071213716695\\
110	4.06517458972563e-06\\
115	0\\
};
\addlegendentry{OLT-R, $\alpha=-2$}

\addplot [color=blue, dotted, line width=1.0pt, mark size=2.0pt, mark=square, mark options={solid, blue}]
  table[row sep=crcr]{%
90	0.133128965665589\\
95	0.0758254971157233\\
100	0.0274549487558807\\
105	0.00210655436006217\\
110	2.63851777826798e-06\\
115	0\\
};
\addlegendentry{OLT-R, $\alpha=-5$}

\end{axis}

\end{tikzpicture}%
			\caption{Steady state $F_{\text{cst}}$ of $F(t)$ with $D=100$.}
			\label{fig:ViolationRate_D100}
		\end{figure}
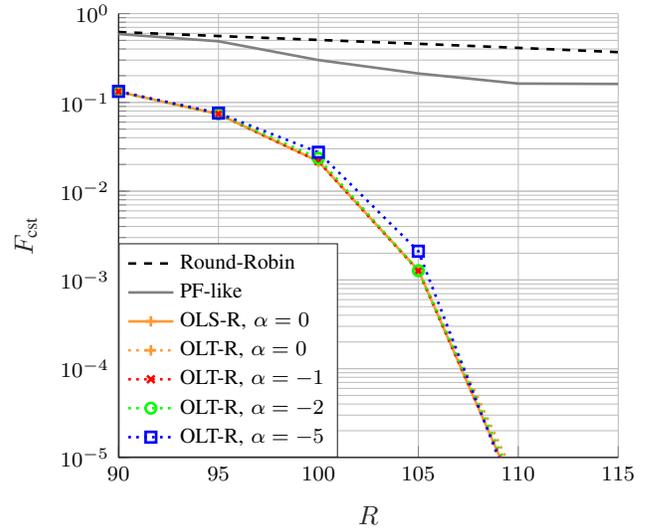

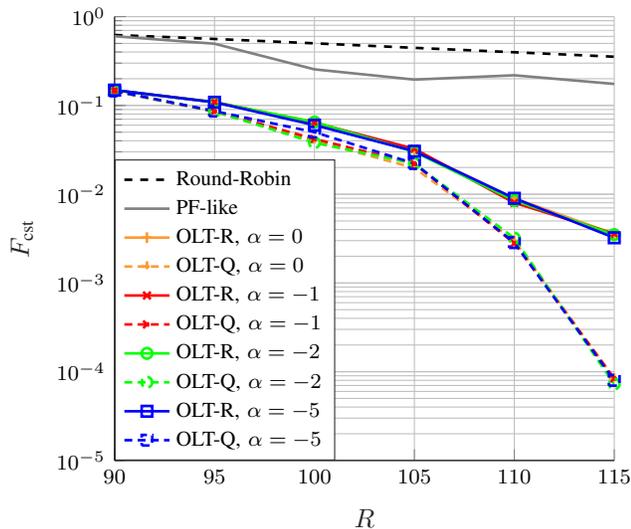
\begin{figure}[!h]
			\centering
%
%
\definecolor{mycolor1}{rgb}{0.00000,0.44700,0.74100}%
\definecolor{mycolor2}{rgb}{0.85000,0.32500,0.09800}%
\definecolor{mycolor3}{rgb}{0.92900,0.69400,0.12500}%
\definecolor{mycolor4}{rgb}{0.49400,0.18400,0.55600}%
\definecolor{mycolor5}{rgb}{0.46600,0.67400,0.18800}%
\definecolor{mycolor6}{rgb}{0.30100,0.74500,0.93300}%
\definecolor{mycolor7}{rgb}{0.63500,0.07800,0.18400}%
\definecolor{orange}{rgb}{1,0.6,0.2}%
\begin{tikzpicture}

\begin{axis}[%
	width=0.75\linewidth,
	height=0.25\textheight,
	font=\footnotesize,
	scale only axis,
	xmin=90,
	xmax=115,
	xlabel style={font=\color{white!15!black}},
	xlabel={$R$},
	ymode=log,
	ymin=0.00001,
	ymax=1,
	yminorticks=true,
	ylabel style={font=\color{white!15!black}},
	ylabel={$F_{\text{cst}}$},
	axis background/.style={fill=white},
	title style={font=\bfseries},
	axis x line*=bottom,
	axis y line*=left,
	xmajorgrids,
	ymajorgrids,
	yminorgrids,
	legend style={at={(0.97,0.03)}, anchor=south east, legend cell align=left, align=left, draw=white!15!black, at={(0,0)}, anchor=south west}
]

\addplot [color=black, dashed, line width=1.0pt]
  table[row sep=crcr]{%
90	0.621631703764468\\
95	0.561091598831226\\
100	0.501115741958674\\
105	0.445757587701156\\
110	0.397464674585225\\
115	0.354091785697323\\
};
\addlegendentry{Round-Robin}

\addplot [color=gray, line width=1.0pt]
  table[row sep=crcr]{%
90	0.60194850855234\\
95	0.496066779097511\\
100	0.25504775647231\\
105	0.195580715148809\\
110	0.218723824058653\\
115	0.174736540048382\\
};
\addlegendentry{PF-like}

\addplot [color=orange, line width=1.0pt, mark=+, mark size=2.0pt]
  table[row sep=crcr]{%
90	0.148193847468249\\
95	0.10925078309069\\
100	0.065\\
105	0.031\\
110	0.009\\
115	0.00356972956755109\\
};
\addlegendentry{OLT-R, $\alpha=0$}

\addplot [color=orange, densely dashed, line width=1.0pt, mark=+, mark size=2.0pt]
  table[row sep=crcr]{%
90	0.143915406171416\\
95	0.0846459441684155\\
100	0.04\\
105	0.019751831510504\\
110	0.003\\
115	8.38106801524272e-05\\
};
\addlegendentry{OLT-Q, $\alpha=0$}

\addplot [color=red, line width=1.0pt, mark=x, mark size=2.0pt]
  table[row sep=crcr]{%
90	0.148504750550628\\
95	0.108810333092258\\
100	0.06365\\
105	0.0325\\
110	0.008\\
115	0.00358838773910007\\
};
\addlegendentry{OLT-R, $\alpha=-1$}

\addplot [color=red, densely dashed, line width=1.0pt, mark=x, mark size=2.0pt]
  table[row sep=crcr]{%
90	0.145615394982499\\
95	0.0858860830725272\\
100	0.042\\
105	0.0221192766703448\\
110	0.0028\\
115	8.38852019436434e-05\\
};
\addlegendentry{OLT-Q, $\alpha=-1$}

\addplot [color=green, line width=1.0pt, mark=o, mark size=2.0pt]
  table[row sep=crcr]{%
90	0.148714250500931\\
95	0.108786540709297\\
100	0.065\\
105	0.03\\
110	0.0085\\
115	0.0035035843350951\\
};
\addlegendentry{OLT-R, $\alpha=-2$}

\addplot [color=green, densely dashed, line width=1.0pt, mark=o, mark size=2.0pt]
  table[row sep=crcr]{%
90	0.145625402966569\\
95	0.0859400018734589\\
100	0.038\\
105	0.0221718936545557\\
110	0.0032\\
115	7.27354827409983e-05\\
};
\addlegendentry{OLT-Q, $\alpha=-2$}

\addplot [color=blue, line width=1.0pt, mark=square, mark size=2.0pt]
  table[row sep=crcr]{%
90	0.14901668752151\\
95	0.108658380781634\\
100	0.06\\
105	0.0305\\
110	0.009\\
115	0.00322916595606319\\
};
\addlegendentry{OLT-R, $\alpha=-5$}

\addplot [color=blue, densely dashed, line width=1.0pt, mark=square, mark size=2.0pt]
  table[row sep=crcr]{%
90	0.145751869490584\\
95	0.0863291683284167\\
100	0.05\\
105	0.0221933052194413\\
110	0.00285\\
115	7.99903113407889e-05\\
};
\addlegendentry{OLT-Q, $\alpha=-5$}

\end{axis}

\end{tikzpicture}%
			\caption{Steady state $F_{\text{cst}}$ of $F(t)$ with $D=90$.}
			\label{fig:ViolationRate_D90}
		\end{figure}

		We show the results for $D=100$ in \figref{fig:ViolationRate_D100} and for $D=90$ in \figref{fig:ViolationRate_D90}. In \figref{fig:ViolationRate_D100}, as the time-lines of the agents are full, OLS-Q behaves exactly as OLS-R and OLT-Q behaves exactly as OLT-R therefore we only display OLS-R and OLT-R. 
		
		First of all, the performance of the Taylor approximation OLT-R are exhibited in comparison with the performance of the exact solver OLS-R in \figref{fig:ViolationRate_D100} for $\alpha=0$. We observe that both are very similar, e.g., for $R=105$, OLS-R and OLT-R both perform nearly with $F_{\text{cst}}=1.2\cdot 10^{-3}$. This confirms that the Taylor approximation is relevant enough to continue only with OLT-R for the other values of $\alpha$ and for $D=100,90$.		
		
		Secondly, we observe that the various values of $\alpha$ for OLT does not bring much different performance. The curves are all very close one to each other therefore we conclude that the value selection for $\alpha$ is sufficiently free, the fairness does not exert a strong constraint on the scheduling performance.
				
		Thirdly, on both figures, we observe that Round-Robin and PF-like similarly perform with poor results, e.g., $F_{\text{cst}}$ does not go below $10^{-1}$ even with great values of $R$. As a matter of fact, Round-Robin does not perform  well because it does not take into account the time-line of the agents and it does not take into account the channel probabilities. In addition, the allocation rule of PF-like does not integrate any form of prediction from the time-line, it only focuses on a past situation of the agents to provide a scheduling decision. Consequently, it integrates somehow the channel error probabilities but it does not take into account the time-lines too which explains the associated bad performance.
		
		Fourthly, we observe a significant gap between the values of $F_{\text{cst}}$ for OLT (R or Q) and PF-like or Round-Robin whatever $D$. This said gap can be exploited in the two following ways:
		\begin{itemize}
			\item We search for the best performance given an application parameter.
			\item We search for the least application constraint that is able to reach a given performance. 
		\end{itemize}
		
		Regarding the first way, for the same application constraint, say $R=100$ and $D=100$, OLT-R exhibits $F_{\text{cst}}=3\cdot 10^{-2}$ whereas PF-like exhibits $F_{\text{cst}}=3\cdot 10^{-1}$, i.e., OLT-R offers a ten times better performance result than PF-like. 
		
		Regarding the second way, to reach the same $F_{\text{cst}}$ performance, e.g., $F_{\text{cst}}=2\cdot 10^{-1}$, OLT-R needs a resilience value of $R=90$ whereas PF-like requires $R=105$, i.e., PF-like enlarges by 17\% the constraint on the application parameters. Furthermore, we observe that PF-like -- and Round-Robin too -- cannot satisfy a performance of $10^{-1}$ or lower whatever the application parameter $R$. Increasing $R$ to even greater values than $115$, the maximum value we considered in the simulations, may indeed not lead to significant better performance for PF-like and Round-Robin. This is not the case of OLT-(R or Q) as the slope of $F_{\text{cst}}$ still increasing (in absolute values) when values of $R$ approach $R=115$. OLT-(R or Q) reaches a $F_{\text{cst}}$ value more than a thousand times less than PF-like at this extreme $R$ value. In other words, playing with $R$ can bring significant benefits for OLT-(R or Q), contrary to PF-like and Round-Robin. This means that the proposed scheduler better integrates the application requirements than the other schedulers.
		
		Comparing now OLT-R with OLT-Q when $D=90$, i.e., when the time-lines get some holes, we observe a performance gap when $R\geq 105$. More precisely, the slope of $F_{\text{cst}}$ for OLT-Q changes with an increase in $R$ whereas the slope of $F_{\text{cst}}$ for OLT-R remains constant. From the application point of view, increasing $R$ means having more application computation capacity, e.g., some extrapolation algorithms in the case of the mobile robot, see \ref{subsec:Example}. This is a constraint regarding the cost of a deployment, therefore, we think it is more beneficial to lower $R$ as much as possible. However, considering the holes in the time-lines requires a slightly more complex scheduler because the computation of $q_k(t)$ is not as easy as the computation of $r_k(t)$. As a matter of fact, we don't need to take into account the time-line to compute $r_k(t)$, only the resilience windows are enough. When computing $q_k(t)$, though, there is a need to couple the knowledge of the resilience windows with the knowledge of the time-lines. In case of some jitter in the application traffic, one needs strong robustness to obtain the exact values of $q_k(t)$. Consequently, selecting either OLT-R or OLT-Q is a question of computation capacity at the application level.
		
		\newpage
\section{Conclusion\label{sec:Conclusion}}
	We designed an application-oriented multi-agent system to better formalize the problem of competition for the access to the radio resource. 
	We jointly considered application-level parameters, characterized mainly by the resilience, and radio environment variables embodied by the channel error probability. 
	We then proposed an on-line scheduler to solve the optimization problem either exactly or approximately with a low-complexity approach. 
	We also introduced a novel way of observing the schedulers' behavior by focusing on an application-level metric instead of focusing on other usual radio-level metrics. 
	From the performance evaluations, we observed that the proposed on-line scheduler design significantly outperform the other schedulers whatever the value of the resilience.
	Moreover, we highlighted that the proposed on-line scheduler allows for more degrees of freedom in the selection of the application parameters to reach an arbitrary performance.
	To conclude, the design of an application-oriented scheduler proved to be a promising method to efficiently integrate application requirements as well as radio parameters.
\bibliographystyle{IEEEtran}
\bibliography{IEEEabrv,biblio}

\end{document}